\documentclass[traditabstract]{aa}

\usepackage{txfonts}
\usepackage{natbib}
   \bibpunct{(}{)}{;}{a}{}{,}
\usepackage{epsfig}
\usepackage{graphicx}
   \DeclareGraphicsExtensions{.eps, .jpg}
\usepackage{color}
\newcommand{\Msun}{M_{\odot}}
\newcommand{\Mdot}{\dot{M}}
\newcommand{\MdotEdd}{\dot{M}_{\rm Edd}}
\newcommand{\LEdd}{L_{\rm Edd}}
\newcommand{\be}{\begin{equation}}                                 
\newcommand{\ee}{\end{equation}}                                   
\newcommand{\bea}{\begin{eqnarray}}                                
\newcommand{\eea}{\end{eqnarray}}                                  
                                                   
 \definecolor{gray}{rgb}{.6,.6,.6}                        
 \definecolor{green}{rgb}{0,.6,0}                         
 \definecolor{red}{rgb}{0.6,0,0}                          
\begin{document}

%=============================================================
\title{
The effect of advection at luminosities close to Eddington: \\ 
The ULX in M31
}

\author
{
Odele Straub \inst{1} \thanks{E-mail: odele.straub@obspm.fr (OS)} 
\and
Chris Done \inst{2} 
\and
Matthew Middleton \inst{3}
}
%=============================================================
\institute{
Laboratoire Univers et Th\'eories, CNRS UMR 8102, Observatoire de Paris, Universit\'e Paris Diderot, 
92190 Meudon, France
\and
Department of Physics, University of Durham, South Road, Durham DH1 3LE, UK
\and
Sterrenkundig Instituut Anton Pannekoek, Universiteit van Amsterdam, 1090 GE Amsterdam, NL
}

%=============================================================
\date{Accepted XXXX. Received XXXX}
%=============================================================
\abstract{
The transient, ultra-luminous X-ray source CXOM31~J004253.1+411422 in the Andromeda galaxy is most likely a 10 solar mass black hole, with super-Eddington luminosity at its peak. The {\it XMM-Newton} spectra taken during the decline then trace luminosities of $0.86-0.27 \LEdd$. These spectra are all dominated by a hot disc component, which roughly follows a constant inner radius track in luminosity and temperature as the source declines. At the highest luminosity the disc structure should change due to advection of radiation through the disc. This advected flux can be partly released at lower radii thus modifying the spectral shape. To study the effect of advection at luminosities close to Eddington we employ a fully relativistic slim disc model, {\tt SLIMBH}, that includes advective cooling and full radiative transfer through the photosphere based on {\sc TLUSTY}. The model also incorporates relativistic photon ray-tracing from the proper location of the disc photosphere rather than the mid-plane as the slim disc is no longer geometrically thin. We find that these new models differ only slightly from the non-advective (standard) {\tt BHSPEC} models even at the highest luminosities considered here. While both discs can fit the highest luminosity data, neither is a very good fit to the {\it lower} luminosities. This could indicate a missing physical process that acts in low luminosity discs and subsides as the disc luminosity approaches the Eddington limit.
%We suggest that this could be due to a decreasing fraction of magnetic pressure support with increasing luminosity, such as seen in models where the magnetic pressure saturates to some fraction of the gas rather than total (gas plus radiation) pressure. 
}
%=============================================================
\authorrunning{O.\,Straub, C. Done \& M. Middleton}
\titlerunning{Advection in spectra of M31 ULX-1}
%=============================================================
\maketitle

%=============================================================
%=============================================================
\section{Introduction}
\label{sec:intro}
%=============================================================
%=============================================================
The most luminous objects in the Universe are powered by accretion of matter onto compact objects like neutron stars and black holes. The energy released depends on the mass and spin of the central black hole as well as the mass accretion rate through the disc, but there is a maximum luminosity for which gravity is able to balance the outward pressure of radiation. Ultra-luminous X-ray sources (ULX) are defined as objects whose bolometric luminosity exceeds this Eddington limit ($L > 10^{39} erg \,\, s^{-1}$). These objects could be powered by normal (sub-Eddington) accretion onto intermediate mass black holes ($M_{BH} \simeq 10^2-10^5 \Msun$). However, there are theoretical problems concerning the formation of such objects \citep{kin+01}, so these are unlikely to form the bulk of the ULX population. Instead, the majority of sources are probably {\it super}-Eddington accreting stellar mass black holes. Such flows come in two flavours, either exceeding the Eddington limit by powering strong outflows \citep{sha+73, lip99, pou+07}, or by advecting the radiation along with the flow (Polish doughnuts: \citet{abr+78,jar+80}; slim discs: \citet{abr+88, sad09}). These processes can occur together \citep{pou+07}, as shown in recent numerical simulations of super-Eddington flows \citep{ohs+09, ohs+11}. In all these cases, radiation is emitted at the local Eddington limit, so that the total source luminosity is $L \sim \LEdd (1 + \ln \Mdot/\MdotEdd)$. Advective accretion discs that can locally exceed the Eddington limit and thus excite winds were proposed by \citet{dot+11}.

The {\tt BHSPEC} models are the best current sub-Eddington accretion flow models for stellar mass black holes. Those calculate the spectrum from full radiative transfer through a disc atmosphere for a Novikov-Thorne (general relativistic) emissivity, including ray-tracing that comprises all general and special relativistic effects \citep{dav+05}. With increasing mass accretion rate, however, the disc's ability to radiate the energy dissipated by viscous stresses ceases. Consequently, the disc overheats and inflates, and the Novikov-Thorne model breaks down. The slim disc is a stable solution for trans-Eddington accretion flows. It is cooled by advection that sweeps some of the emitted energy along with the flow as photons are trapped in the optically thick disc. The photons can be released again at lower radii as the material accelerates towards the black hole. The {\tt SLIMBH} models, a modification of {\tt BHSPEC}, incorporate the effects of advection, such as the increasing height of the photosphere and the shift of the inner disc edge towards smaller radii \citep{sad+11}. {\tt SLIMBH} models were recently fitted to spectra from the accreting black hole in LMC X-3. However, the differences between standard thin disc models and {\tt SLIMBH} are not large there although this object reaches $\sim 0.6 \LEdd$ \citep{str+11}.

Here instead we fit these new models to the higher Eddington fraction flows seen in ULX. In particular, we use the XMM-Newton spectrum from the transient ULX CXOM31 J004253.1+411422 (hereafter M31 ULX-1). The advantage of this object is that it showed a steady exponential decline in luminosity in five XMM-Newton spectra from $1.6 - 0.8 \times 10^{39} erg \,\, s^{-1}$ after its discovery at $5 \times 10^{39} erg \,\, s^{-1}$ in Chandra imaging data. Such exponential decays with this time scale are well known from transient low mass X-ray binary black holes in our Galaxy, making it most probable that this ULX is also a $\sim10 \Msun$ black hole \citep{mid+12, kau+12}. This mass would then imply that the luminosity sampled by the XMM-Newton spectra corresponds to $0.86-0.27 \LEdd$, considerably higher than seen in LMC X-3. Thus it provides an ideal testing ground for the slim disc models, especially as the Galactic absorption column is fairly low, allowing the broad band disc shape to be seen down to low energies.

The paper is structured as follows. The origin and analysis of our data is specified in section~\ref{sec:data}, the modelling of the ULX spectra is described in section~\ref{sec:modelling} and the results are discussed in section~\ref{sec:discussion}.

% ================================================
% ================================================
\section{X-ray data: reduction and analysis}
\label{sec:data}
% ================================================
% ================================================
To ensure consistency, we followed the procedure of data reduction detailed in \citet{mid+12} and extracted the event files using {\sc sas v10} and filtered the products for standard patterns ($<$=12 for MOS, $<$= 4 for PN) and flags (=0 for spectral and timing products). The full field hard (10--15~keV) count rate was used to create good time intervals excluding contamination by soft proton flares \citep[see Table 1 of][]{mid+12}. These were then used to extract spectral (including responses) and timing products from a circular, 35'' radius source, with background regions taken from the same chip, avoiding other sources in the field. 

The first observation was slightly piled up (seen using the tool, {\sc epatplot}), so the inner 5'' centroid centred on the source was removed in the PN. The MOS is more affected by pileup, so was discarded for this observation.

% ================================================
% ================================================
\section{Modelling thermal ULX spectra}
\label{sec:modelling}
% ================================================
% ================================================
The trans-Eddington luminosity regime of ULX spectra falls in the domain of slim disc accretion. {\tt SLIMBH} is a fully relativistic slim disc model based on \citet{sad+11} that accounts for effects connected to high mass accretion rates such as advection of radiation, the relocation of the inner disc edge towards radii smaller than the innermost stable circular orbit (ISCO), and the proper, non-equatorial location of the disc photosphere. For the vertical radiative transfer {\tt SLIMBH} directly incorporates the {\sc TLUSTY} grid of disc local annuli spectra \citep{hub+95} which are then integrated and ray-traced through the relativistic spacetime. It is thus constructed in the same manner as {\tt BHSPEC} \citep{dav+06}. The difference between the two models is the underlying accretion disc: While {\tt BHSPEC} is based on a standard thin \citet{nov+73} disc, {\tt SLIMBH} uses the slim disc \citep{sad+11}. A comparison of these two models offers valuable insight on the effect of advection (see section~\ref{sec:SLIMBH_vs_BHSPEC}).

\citet{mid+12} fit the five XMM-Newton spectra of M31 ULX-1 simultaneously with {\tt TBABS} $\times$ {\tt BHSPEC} using the neutral hydrogen column density $N_H = 6.7 \times 10^{20} \,\, cm^{-2}$ \citep[after][]{dic+90}, distance $D = 780$ kpc \citep{vil+10,tan+10} and viscosity parameter $\alpha = 0.01$. With black hole mass fixed to the provisional value of $M = 10 \Msun$ they derived a best fit inclination of $i = 30^{\circ}$ and dimensionless spin parameter $a_* = 0.36^{+0.10}_{-0.11}$. The fit is fairly good, with $\chi^2_\nu=3169.4/2801$, though the spectra are significantly better fit with a {\it phenomenological} model of {\tt TBABS $\times$ (DISKPBB+COMPTT)} ($\chi^2_\nu=2752.2/2777$).

\begin{figure}[ht]
\includegraphics[width=0.5\textwidth]{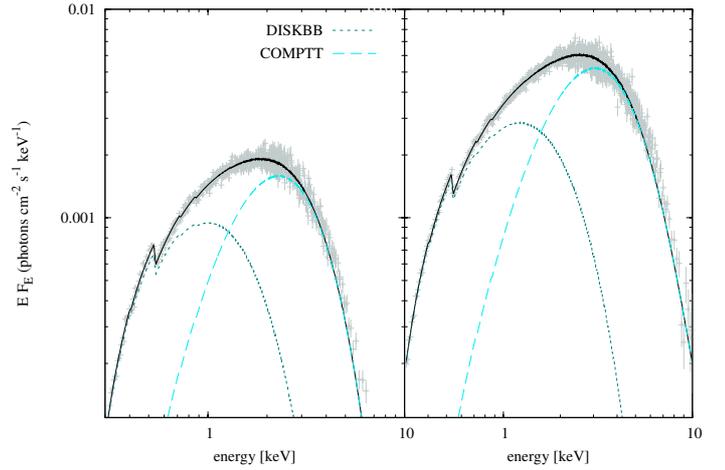}
\caption{Synthetic slim disc spectra produced with {\tt SLIMBH} and fitted to {\tt DISKBB+COMPTT}. The total model (solid black) is given by the sum of the disc component and the Compton component. {\it Left}: Low luminosity case, $L = 0.27 \LEdd$. {\it Right}: High luminosity case, $L = 0.86 \LEdd$.}
\label{fig:dontdoit}
\end{figure}

A phenomenological model does not necessarily represent the real circumstances. To illustrate this we use {\tt SLIMBH} to produce two synthetic datasets that represent ideal slim discs at low and high luminosity. We then fit these simulated data with the phenomenological {\tt DISKPBB+COMPTT} model. From figure~\ref{fig:dontdoit} it seems to be the case that both, the low and the high luminosity spectrum exhibit a notable amount of Comptonisation which, by construction, is not present. The spectra of M31 ULX-1 show no evidence of Comptonisation. The fact that the phenomenological model gives a statistically better fit tells us only that there are enough free parameters to match the data. In the following we focus on physical models. We use, however, the phenomenological ones to provide us with the reference spectral shape of an ``ideal'' disc model.

% ================================================
% FIGURE -- 1
% ================================================
% ================================================

% ================================================
% TABLE 1 -- SLIM
% ================================================
\begin{table}[ht]
\centering
\caption{Slim disc model: best fitting spectral parameters.}
\begin{tabular}{lccccc}
\hline \hline 
 & \multicolumn{5}{c}{{\tt TBABS} * {\tt SLIMBH}}  \\ 
\hline \hline 
Observation & 1 & 2 & 3 & 4 & 5 \\ 
\hline 
$N_{\rm H}$ [$cm^{-2}$]  & \multicolumn{5}{c}{$ 6.7 \times 10^{\,20} $}  \\ [5pt]
$M_{\rm BH}$ [$\Msun$] & \multicolumn{5}{c}{10} \\ [5pt]
$a_*$ &  \multicolumn{5}{c}{$0.36^{+0.02}_{-0.02}$} \\ [6pt]
$L/\LEdd$ & $0.86^{+0.01}_{-0.01}$ & $0.59^{+0.01}_{-0.01}$ & $0.43^{+0.01}_{-0.01}$ & $0.34^{+0.01}_{-0.01}$ & $0.27^{+0.01}_{-0.01}$  \\[5pt]
$\alpha$ & \multicolumn{5}{c}{0.01} \\ [5pt]
i [$ ^{\circ}$] & \multicolumn{5}{c}{30}  \\ [5pt]
D [kpc] & \multicolumn{5}{c}{780}  \\ [5pt]
$\chi/dof$ ($\chi^2_\nu$) & \multicolumn{5}{c}{3132.2/2801 (1.12)}  \\ [5pt]
\hline
%
%$N_{\rm H}$ [$cm^{-2}$]  & \multicolumn{5}{c}{$ 6.7 \times 10^{\,20} $}  \\ [5pt]
%
%$M_{\rm BH}$ [$\Msun$] & \multicolumn{5}{c}{10} \\ [5pt]
%
%$a_*$ &  $0.33^{+0.03}_{-0.02}$ & $0.40^{+0.02}_{-0.03}$ & $0.42^{+0.04}_{-0.04}$ & $0.34^{+0.04}_{-0.04}$ & $0.34^{+0.04}_{-0.03}$\\ [6pt]
%
%$L/\LEdd$ & $0.85^{+0.01}_{-0.01}$ & $0.60^{+0.01}_{-0.01}$ & $0.44^{+0.01}_{-0.01}$ & $0.33^{+0.01}_{-0.01}$ & $0.27^{+0.01}_{-0.01}$ \\[5pt]
%
%$\alpha$ & \multicolumn{5}{c}{0.01} \\ [5pt]
%
%i [$ ^{\circ}$] & \multicolumn{5}{c}{30}  \\ [5pt]
%
%D [kpc] & \multicolumn{5}{c}{780}  \\ [5pt]
%
%$\chi/dof$ ($\chi^2_\nu$) &  \multicolumn{5}{c}{3118.1/2797 (1.12)} \\ 
%\hline
\label{tab:slim_fit}
\end{tabular}
\end{table}
% ================================================
% ================================================

We now replace the {\tt BHSPEC} model with {\tt SLIMBH} using the same parameter specifications as above. The objective is to see if an advective disc can better reproduce the spectral shape, and provide a {\it physical} model which fits the data as well as the phenomenological one. Advection should become increasingly important after $L \sim 0.1 \LEdd$, so the differences between {\tt BHSPEC} and {\tt SLIMBH} could become notable. We find that the slim disc models indeed fit the data slightly better, but that the improvement is only small, with $\chi^2_\nu=3132.2/2801$ for $\alpha = 0.01$ and $\chi^2_\nu=3131.6/2801$ for $\alpha = 0.1$, and the respective black hole spin $a_* = 0.36$ and $a_* = 0.26$. 

%The $\alpha$-viscosity prescription is an ad hoc assumption used to describe the stress-pressure relation in standard and slim disc models. Studies show that $\alpha = 0.1$ is consistent with data from cataclysmic variables \citep{kin+07}, but inconsistent with data from X-ray binaries at moderate luminosities \citep[e.g.,][]{don+08, str+11}. 

% ================================================
% FIGURE -- 2
% ================================================
\begin{figure*}[t!]
\includegraphics[width=\textwidth]{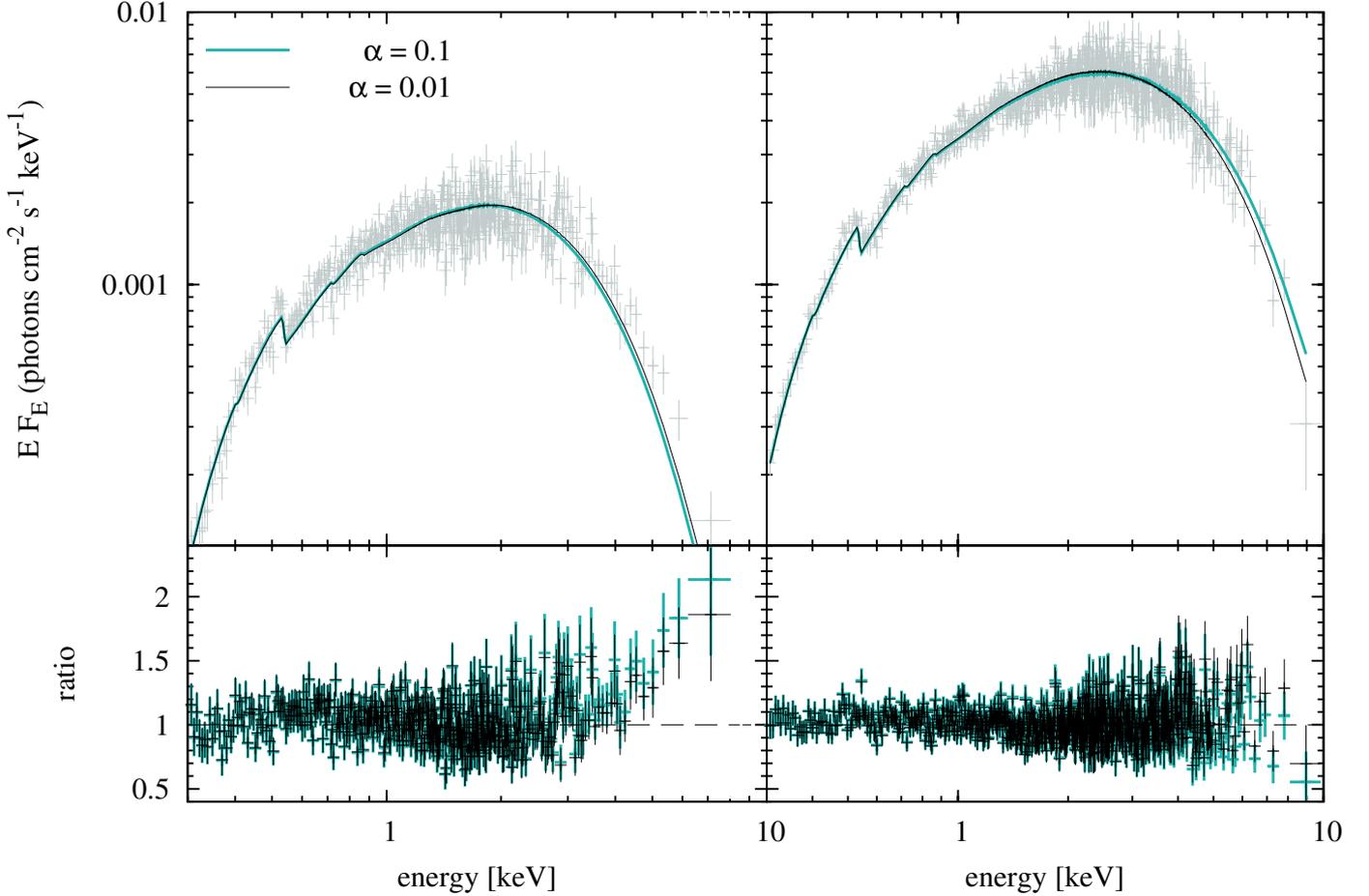}
\caption{X-ray spectra of M31 ULX-1, modelled with {\tt SLIMBH}. {\it Left}: Spectra for $\alpha = 0.1$ and 0.01 with the respective data/model ratio for the lowest ($= 0.27 \LEdd$) luminosity data set. {\it Right}: Spectra for $\alpha = 0.1$ and 0.01 with the respective data/model ratio for the highest ($= 0.86 \LEdd$) luminosity data set.}
\label{fig:high-low}
\end{figure*}
% ================================================

A closer inspection of the spectral residuals shown in \citet{mid+12} explains this result: The residuals between {\tt BHSPEC} and the data are largest at {\it lowest} luminosities, with the data being increasingly well fit by the non-advective disc models at higher luminosities. This is opposite to the behaviour expected if advection becomes increasingly important, limiting the impact of the spectral changes from {\tt SLIMBH}. Similar residuals appear when fitting the M31 ULX-1 spectra with {\tt SLIMBH} as shown for the lowest and highest luminosity dataset in figure~\ref{fig:high-low}. This behaviour is not affected by the value of the $\alpha$-parameter nor by the tied spin. We note, however, that the lower $\alpha$ value fits the lowest and highest luminosity data marginally better. Models with a free spin parameter fit the data equally well as those with tied spin and exhibit the same residuals (see the residuals in figure~\ref{fig:high-low}). The free spin then assumes values between $a_* = 0.33$ and 0.42 (with errors between 5-10\%), where the highest are estimated for $L = 0.4 - 0.6 \LEdd$ and the lowest for $L = 0.86 \LEdd$. The measured tied spin lies within the error bars of all free spin values. We explore the differences between {\tt BHSPEC}, {\tt SLIMBH} and the data in more detail below.

% ================================================
% ================================================
\subsection{Effect of advection: The difference between {\tt SLIMBH} and {\tt BHSPEC}}
\label{sec:SLIMBH_vs_BHSPEC}
\begin{figure*}[ht!]
\includegraphics[width=\textwidth]{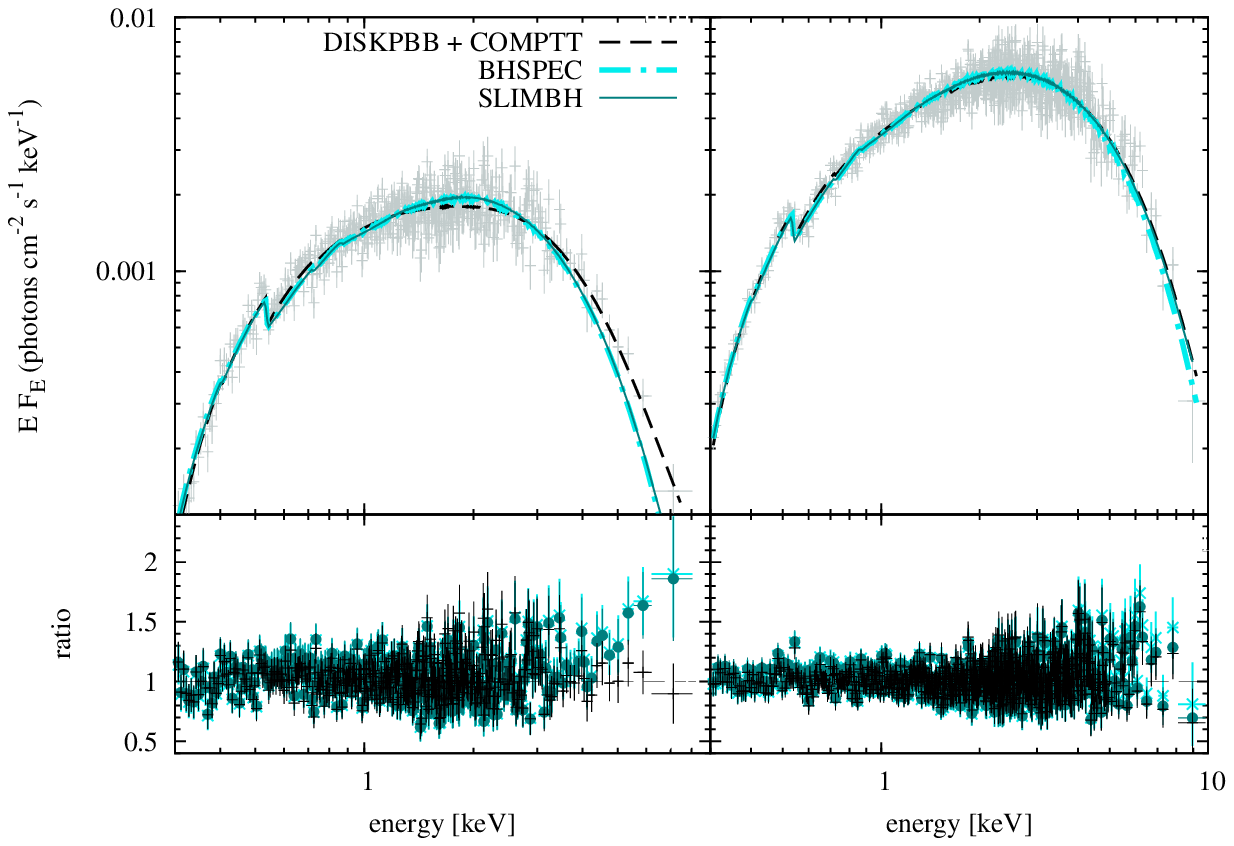}
\caption{{\tt SLIMBH} and {\tt BHSPEC} versus the phenomenological 2-component model. The black hole mass, $\alpha$ and inclination are fixed at $10\Msun$, 0.01 and $30^\circ$, respectively, and the black hole spin is tied, $a_* = 0.36$. All three models represent best fits. The data are well described by the physical disc models at the highest luminosity, but are significantly broader at lower luminosity. {\it Left}: Low luminosity disc ($L = 0.27 \LEdd$). {\it Left}: High luminosity disc ($L = 0.86 \LEdd$).}
\label{fig:comp_spectra}
\end{figure*}

\begin{figure}[t]
\includegraphics[width=0.5\textwidth]{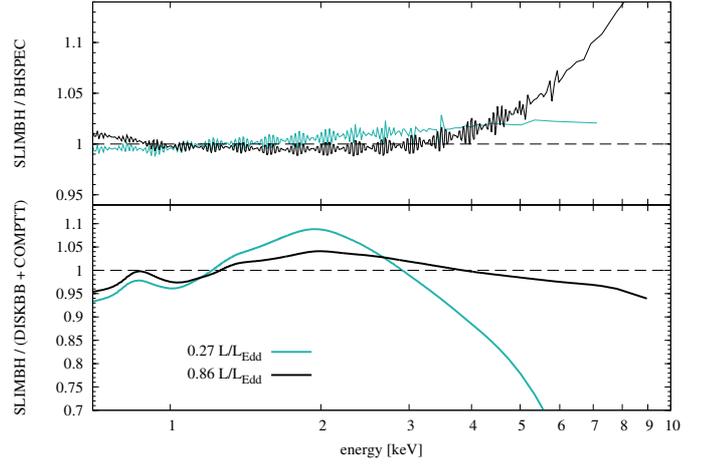}
\caption{The high and low luminosity model ratios {\tt SLIMBH} vs {\tt BHSPEC} (upper panel) and {\tt SLIMBH} vs {\tt DISKPBB+COMPTT} (lower panel) for the same spectra as in figure~\ref{fig:comp_spectra}. Note the different scales on the y-axes.}
\label{fig:comp_ratio}
\end{figure}

% ================================================
% ================================================
Figure~\ref{fig:comp_spectra} shows the best fit model spectra of {\tt SLIMBH} and {\tt BHSPEC}. Their ratio for identical mass ($10 \Msun$), spin ($a_* = 0.36$), inclination ($30^\circ$), viscosity parameter ($\alpha = 0.01$) and luminosity ($= 0.27/0.86 \LEdd$) is presented in the top panel of figure~\ref{fig:comp_ratio}. The figure shows that the low luminosity model ratio is not exactly one. This inequality of the models can be attributed to the ray-tracing from the proper photosphere location in the case of {\tt SLIMBH} and the fact that already at $L = 0.27 \LEdd$ the disc height is not thin anymore. The high luminosity model ratio confirms that the slim disc is very slightly softer below the peak as radiation is transported inwards at these radii rather than being emitted. Some fraction of this advected radiation can then be released at lower radii, so there is additional radiation in the {\tt SLIMBH} models at the highest energies i.e. above the peak. Thus the {\tt SLIMBH} models are broader and not so peaked as the corresponding {\tt BHSPEC} models, but the difference is less than a few per cent for $0.86 \LEdd$, and there is even less difference between {\tt SLIMBH} and {\tt BHSPEC} at lower luminosities. Advection only makes a more substantial difference to the emitted spectrum for super-Eddington luminosities.  

% ================================================
% FIGURE -- 3
% ================================================
% ================================================

However, there is a significant change in how well the models describe the data. The black dashed lines in figure~\ref{fig:comp_spectra} represent the best fit phenomenological {\tt DISKPBB+COMPTT} model for the highest/lowest luminosity data of M31 ULX-1. The high luminosity dataset is well described by either disc model. By contrast, the low luminosity dataset is clearly broader than the corresponding {\tt BHSPEC} and {\tt SLIMBH} models. The discrepancies between physical and phenomenological models are illustrated in the bottom panel of figure~\ref{fig:comp_ratio} where the model ratio {\tt SLIMBH} vs {\tt DISKPBB+COMPTT} for the highest (black) and lowest (cyan) luminosity is given. We stress that adding a Compton component to the slim disc is not a good way to model the spectra at hand. {\tt SLIMBB+COMPTT} fits the data with a fairly steep powerlaw that extends below the peak of the disc flux, i.e., below its seed photons. This, however, is not a physical scenario of Compton scattering.

% ================================================
% FIGURE -- 4
% ================================================
% ================================================

Thus, including advection does not change the conclusion of \citet{mid+12} that the data (especially at lower luminosity) are better fit by phenomenological two component models rather than physical disc models. This shows that there is something missing from our best current disc models. We stress that the level is only 5-10 per cent, but this is significant given the quality of data now available. Moreover, the mismatch between models and data in low luminosity spectrum is strongest in the high energy band that is crucial for the estimation of the black hole spin. The missing physics is not advection, nor is it any of the other spectral distortions which might become more important as the luminosity increases, such as winds, or bulk motion turbulence in the disc, as the data become {\it better} fit by our disc models as the luminosity approaches Eddington.

% ================================================
% ================================================
\section{Discussion and Conclusions}
\label{sec:discussion}
% ================================================
% ================================================
It has long been assumed that in the trans-Eddington luminosity regime two effects could become increasingly important for accretion disc models, (i) advection \citep{abr+88,min+00} and/or (ii) outflows \citep{sha+73, pou+07, ohs+09}. In this paper we pin down the effect of advection by fitting ULX spectra with slim discs. We find that advection typically changes the best fitting spectral shape by removing photons in the intermediate energy range where the spectral peak is located and by releasing more photons at high energies (see in Figure~\ref{fig:comp_ratio}, upper panel). This is the characteristic signature of slim discs where photons are trapped in the accretion flow, accreted inward and released closer to the black hole and even inside the plunging region (see \citet{abr+10} for a discussion of the inner edge of slim discs and \citet{zhu+12} for the relevance of radiation emitted from inside the plunging region).

However, these changes make very little difference to the emitted spectrum for $L < \LEdd$, and we show here that this cannot be the reason for the mismatch between data and models seen in moderate luminosity ($0.1-0.5 \LEdd$) spectra from both this ULX and from galactic black hole binaries \citep{kol+11,str+13}. Real accretion discs at these moderate luminosities have a spectrum which is subtly broader than is currently expected from the best disc models. This additional broadening is most marked at lower mass accretion rates and gets weaker as the disc luminosity increases. This is exactly opposite to the luminosity behaviour expected from either winds, advection, bulk turbulence and/or inhomogeneities in the disc. 

Instead, it could be connected to the (low level) coronal emission. We cannot constrain a weak high energy tail in our data, but this could potentially illuminate the disc and hence change the structure of the photosphere. However, we note that the most disc dominated GX339-4 spectrum in \citet{kol+11}, where the tail carries less than a few per cent of the total bolometric flux, shows this additional broadening. This is a very low level of illumination to cause a noticeable change in disc structure. Self-illumination of the inner disc by light-bending around the black hole may be a larger effect when the coronal emission is low \citep[e.g.,][]{min+04}. However, this should involve a constant fraction of the emission, rather than give a decreasing fraction as $L/\LEdd$ increases. 

Here we speculate that the broadened spectral shape at low luminosity may actually be a signature of magnetic pressure support in the disc. This will increase the scale height of the disc over and above that expected from a standard disc, so decreasing its density and hence its true absorption opacity. Electron scattering then becomes more important, broadening the spectrum \citep{dav+09}. The extent of this change in disc structure depends on the ratio of magnetic pressure to total (radiation plus gas plus magnetic) pressure. This could explain the observed trend in the data if the magnetic pressure saturates to some fraction of the gas pressure rather than total pressure, as it would decrease in importance as $L/\LEdd$ increases. 

Other observations also strongly indicate that there is a change in behaviour in magnetic stress scaling between the gas pressure and radiation pressure dominated regimes. The rapid rise to outburst, which occurs (mostly) in the gas pressure dominated regime, requires very efficient transport of angular momentum as parametrised by a Shakura-Sunyaev $\alpha$ viscosity of $\sim 0.1$ \citep{dub+99}. However, if this efficient transport is maintained into the radiation pressure dominated regime then the disc becomes effectively optically thin, and its colour temperature correction increases markedly. This is in sharp contrast to the data, where the observed constancy of the colour temperature correction requires $\alpha<0.01$ at high $L/\LEdd$, where the disc is in the radiation pressure dominated regime \citep{don+08, str+11}. Thus the data from the ULX reported here, and previous work on the disc spectra seen from black hole binaries both support a change in magnetic stress scaling with total pressure. Current numerical simulations do not show this behaviour \citep{hir+09}, but these are at the limit of computational capabilities. Future research on how the self generated magnetic field from the MRI scales with pressure could resolve this issue.

% ================================================
% ================================================
\section*{Acknowledgments}
% ================================================
% ================================================
OS thanks the Department of Physics at the University of Durham, and in particular Martin Ward for their hospitality. CD acknowledges illuminating conversations with Shane Davis on magnetic pressure support for the disc.

\bibliographystyle{aa}
\bibliography{advection_in_ULX}

\begin{thebibliography}{32}
\expandafter\ifx\csname natexlab\endcsname\relax\def\natexlab#1{#1}\fi

\bibitem[{{Abramowicz} {et~al.}(1978){Abramowicz}, {Jaroszynski}, \&
  {Sikora}}]{abr+78}
{Abramowicz}, M., {Jaroszynski}, M., \& {Sikora}, M. 1978, \aap, 63, 221

\bibitem[{{Abramowicz} {et~al.}(1988){Abramowicz}, {Czerny}, {Lasota}, \&
  {Szuszkiewicz}}]{abr+88}
{Abramowicz}, M.~A., {Czerny}, B., {Lasota}, J.~P., \& {Szuszkiewicz}, E. 1988,
  \apj, 332, 646

\bibitem[{{Abramowicz} {et~al.}(2010){Abramowicz}, {Jaroszy{\'n}ski}, {Kato},
  {Lasota}, {R{\'o}{\.z}a{\'n}ska}, \& {S{\c a}dowski}}]{abr+10}
{Abramowicz}, M.~A., {Jaroszy{\'n}ski}, M., {Kato}, S., {et~al.} 2010, \aap,
  521, A15

\bibitem[{{Davis} {et~al.}(2009){Davis}, {Blaes}, {Hirose}, \&
  {Krolik}}]{dav+09}
{Davis}, S.~W., {Blaes}, O.~M., {Hirose}, S., \& {Krolik}, J.~H. 2009, \apj,
  703, 569

\bibitem[{{Davis} {et~al.}(2005){Davis}, {Blaes}, {Hubeny}, \&
  {Turner}}]{dav+05}
{Davis}, S.~W., {Blaes}, O.~M., {Hubeny}, I., \& {Turner}, N.~J. 2005, \apj,
  621, 372

\bibitem[{{Davis} \& {Hubeny}(2006)}]{dav+06}
{Davis}, S.~W. \& {Hubeny}, I. 2006, \apjs, 164, 530

\bibitem[{{Dickey} \& {Lockman}(1990)}]{dic+90}
{Dickey}, J.~M. \& {Lockman}, F.~J. 1990, \araa, 28, 215

\bibitem[{{Done} \& {Davis}(2008)}]{don+08}
{Done}, C. \& {Davis}, S.~W. 2008, \apj, 683, 389

\bibitem[{{Dotan} \& {Shaviv}(2011)}]{dot+11}
{Dotan}, C. \& {Shaviv}, N.~J. 2011, \mnras, 413, 1623

\bibitem[{{Dubus} {et~al.}(1999){Dubus}, {Lasota}, {Hameury}, \&
  {Charles}}]{dub+99}
{Dubus}, G., {Lasota}, J.-P., {Hameury}, J.-M., \& {Charles}, P. 1999, \mnras,
  303, 139

\bibitem[{{Hirose} {et~al.}(2009){Hirose}, {Blaes}, \& {Krolik}}]{hir+09}
{Hirose}, S., {Blaes}, O., \& {Krolik}, J.~H. 2009, \apj, 704, 781

\bibitem[{{Hubeny} \& {Lanz}(1995)}]{hub+95}
{Hubeny}, I. \& {Lanz}, T. 1995, \apj, 439, 875

\bibitem[{{Jaroszynski} {et~al.}(1980){Jaroszynski}, {Abramowicz}, \&
  {Paczynski}}]{jar+80}
{Jaroszynski}, M., {Abramowicz}, M.~A., \& {Paczynski}, B. 1980, \actaa, 30, 1

\bibitem[{{Kaur} {et~al.}(2012){Kaur}, {Henze}, {Haberl}, {Pietsch}, {Greiner},
  {Rau}, {Hartmann}, {Sala}, \& {Hernanz}}]{kau+12}
{Kaur}, A., {Henze}, M., {Haberl}, F., {et~al.} 2012, \aap, 538, A49

\bibitem[{{King} {et~al.}(2001){King}, {Davies}, {Ward}, {Fabbiano}, \&
  {Elvis}}]{kin+01}
{King}, A.~R., {Davies}, M.~B., {Ward}, M.~J., {Fabbiano}, G., \& {Elvis}, M.
  2001, \apjl, 552, L109

\bibitem[{{Kolehmainen} {et~al.}(2011){Kolehmainen}, {Done}, \& {D{\'{\i}}az
  Trigo}}]{kol+11}
{Kolehmainen}, M., {Done}, C., \& {D{\'{\i}}az Trigo}, M. 2011, \mnras, 416,
  311

\bibitem[{{Lipunova}(1999)}]{lip99}
{Lipunova}, G.~V. 1999, Astronomy Letters, 25, 508

\bibitem[{{Middleton} {et~al.}(2012){Middleton}, {Sutton}, {Roberts},
  {Jackson}, \& {Done}}]{mid+12}
{Middleton}, M.~J., {Sutton}, A.~D., {Roberts}, T.~P., {Jackson}, F.~E., \&
  {Done}, C. 2012, \mnras, 420, 2969

\bibitem[{{Mineshige} {et~al.}(2000){Mineshige}, {Kawaguchi}, {Takeuchi}, \&
  {Hayashida}}]{min+00}
{Mineshige}, S., {Kawaguchi}, T., {Takeuchi}, M., \& {Hayashida}, K. 2000,
  \pasj, 52, 499

\bibitem[{{Miniutti} \& {Fabian}(2004)}]{min+04}
{Miniutti}, G. \& {Fabian}, A.~C. 2004, \mnras, 349, 1435

\bibitem[{{Novikov} \& {Thorne}(1973)}]{nov+73}
{Novikov}, I.~D. \& {Thorne}, K.~S. 1973, in Black Holes (Les Astres Occlus),
  343--450

\bibitem[{{Ohsuga} \& {Mineshige}(2011)}]{ohs+11}
{Ohsuga}, K. \& {Mineshige}, S. 2011, \apj, 736, 2

\bibitem[{{Ohsuga} {et~al.}(2009){Ohsuga}, {Mineshige}, {Mori}, \&
  {Kato}}]{ohs+09}
{Ohsuga}, K., {Mineshige}, S., {Mori}, M., \& {Kato}, Y. 2009, \pasj, 61, L7+

\bibitem[{{Poutanen} {et~al.}(2007){Poutanen}, {Lipunova}, {Fabrika},
  {Butkevich}, \& {Abolmasov}}]{pou+07}
{Poutanen}, J., {Lipunova}, G., {Fabrika}, S., {Butkevich}, A.~G., \&
  {Abolmasov}, P. 2007, \mnras, 377, 1187

\bibitem[{{S{\c a}dowski}(2009)}]{sad09}
{S{\c a}dowski}, A. 2009, \apjs, 183, 171

\bibitem[{{S{\c a}dowski} {et~al.}(2011){S{\c a}dowski}, {Abramowicz}, {Bursa},
  {Klu{\'z}niak}, {Lasota}, \& {R{\'o}{\.z}a{\'n}ska}}]{sad+11}
{S{\c a}dowski}, A., {Abramowicz}, M., {Bursa}, M., {et~al.} 2011, \aap, 527,
  A17

\bibitem[{{Shakura} \& {Sunyaev}(1973)}]{sha+73}
{Shakura}, N.~I. \& {Sunyaev}, R.~A. 1973, \aap, 24, 337

\bibitem[{{Straub} {et~al.}(2011){Straub}, {Bursa}, {S{\c a}dowski}, {Steiner},
  {Abramowicz}, {Klu{\'z}niak}, {McClintock}, {Narayan}, \&
  {Remillard}}]{str+11}
{Straub}, O., {Bursa}, M., {S{\c a}dowski}, A., {et~al.} 2011, \aap, 533, A67

\bibitem[{{Straub} \& {Ghasemnezhad}(2013)}]{str+13}
{Straub}, O. \& {Ghasemnezhad}, M. 2013, \textit{in prep.}

\bibitem[{{Tanaka} {et~al.}(2010){Tanaka}, {Chiba}, {Komiyama}, {Guhathakurta},
  {Kalirai}, \& {Iye}}]{tan+10}
{Tanaka}, M., {Chiba}, M., {Komiyama}, Y., {et~al.} 2010, \apj, 708, 1168

\bibitem[{{Vilardell} {et~al.}(2010){Vilardell}, {Ribas}, {Jordi},
  {Fitzpatrick}, \& {Guinan}}]{vil+10}
{Vilardell}, F., {Ribas}, I., {Jordi}, C., {Fitzpatrick}, E.~L., \& {Guinan},
  E.~F. 2010, \aap, 509, A70

\bibitem[{{Zhu} {et~al.}(2012){Zhu}, {Davis}, {Narayan}, {Kulkarni}, {Penna},
  \& {McClintock}}]{zhu+12}
{Zhu}, Y., {Davis}, S.~W., {Narayan}, R., {et~al.} 2012, \mnras, 424, 2504

\end{thebibliography}

\end{document}